\documentclass[10pt,conference]{IEEEtran}

\usepackage{xspace}
\usepackage{graphicx}
\usepackage{textcomp}
\usepackage{xcolor}
\usepackage{siunitx}
\usepackage{xurl} % This is used to be able to break long \url{...} at arbitrary location
\usepackage{hyperref}
\hypersetup{breaklinks=true} % This is used to be able to break long \url{...} at arbitrary location

\AtBeginDocument{%
  }

\begin{document}

\title{Using Azure Quantum Resource Estimator for Assessing Performance of Fault Tolerant Quantum Computation}

\author{
  \IEEEauthorblockN{Wim van Dam}
  \IEEEauthorblockA{wimvandam@microsoft.com \\ Microsoft Quantum, United States}
% }
\and 
% \author{
  \IEEEauthorblockN{Mariia Mykhailova}
  \IEEEauthorblockA{mamykhai@microsoft.com \\ Microsoft Quantum, United States}
% }
\and 
% \author{
  \IEEEauthorblockN{Mathias Soeken}
  \IEEEauthorblockA{mathias.soeken@microsoft.com \\ Microsoft Quantum, Switzerland}
}

\newcommand{\AQRE}{\textsc{aqre}\xspace}
\newcommand{\AQDK}{\textsc{aqdk}\xspace}

\maketitle

\begin{abstract}
The resource estimation tools provided by Azure Quantum and Azure Quantum Development Kit are described. 
Using these tools one can automatically evaluate the logical and physical resources required to run algorithms on fault-tolerant quantum computers. 
An example is given of obtaining resource estimates for quantum fault-tolerant implementations of three different multiplication algorithms.

\end{abstract}

% \tableofcontents

\section{Introduction}

Quantum computing promises to solve scientifically and commercially important problems that are intractable for classical computation.
Delivering on this promise requires building quantum supercomputers---large-scale, error corrected quantum computers with reliable, logical qubits, and fault tolerant operations. 
To quantify the technical requirements for building such quantum supercomputers, we need to be able to a priori estimate the resource requirements of implementing the relevant quantum algorithms under realistic assumptions.
Doing so will achieve at least two goals. 
On the physical side, resource estimation gives us necessary and sufficient conditions that quantum machines need to meet to be considered practical. 
On the logical side,  resource estimation clarifies which algorithms truly give a quantum advantage over classical computation, and which quantum algorithms do not meet this bar. 

In this paper, we describe Azure Quantum Resource Estimator---an open-source tool developed by Microsoft's Azure Quantum team that estimates the logical and physical resources required to run quantum algorithms on a fault-tolerant quantum computer. 
In Section~\ref{sec:algorithm} we describe the algorithm that the tool uses for its resource estimation. 
Section~\ref{sec:AQRE} walks through the algorithmic inputs, the resource estimation parameters, and the output data of the the tool. 
We conclude with an example case study that compares three quantum algorithms for large integer multiplication. 
The scientific background of this tool is described in an earlier technical paper~\cite{Beverland2022}.

\section{Background}

To measure progress toward a fault-tolerant quantum supercomputer, we use the following three \emph{quantum computing implementation levels}, which are based on the computational capabilities of the quantum processor~\cite{rQOPS-Nayak}. 

\subsubsection*{Level~1: Foundational quantum computing}
Today’s Noisy Intermediate Scale Quantum (NISQ) computers use noisy physical qubits as their computational units~\cite{NISQ}. 
While such quantum systems are valuable from a scientific point of view, they are not be able to solve commercially relevant problems in a manner that is more efficient than what we can do with classical computers. 
The reason for this inability lies with the experimental fact that physical noise rates will prevent NISQ circuits from computing beyond, at most, a few thousand quantum gates. 

\subsubsection*{Level~2: Resilient quantum computing}
We speak of resilient quantum computing when we are able to implement a logical qubit whose reliability exceeds the physical error rates of its components. 
Such logical qubits use quantum error correction on a multitude of physical qubits per logical qubit. 

\subsubsection*{Level~3: Scalable quantum computing}
With enough reliable qubits and an appropriate logical clock speed we reach the third level of quantum computing at scale.
It is at this level that we have a quantum supercomputer that gives us a quantum advantage over classical computers for commercially valuable problems.

To outperform classical computation for practical applications, we need the capability to reliably execute algorithms with $\num{e12}$ quantum gates or more~\cite{Beverland2022}.
Moreover, for the quantum computation to provide the solution within a practical amount of time, say within \num{e6} seconds, we also have to take into account the logical clock speed of the quantum computers, which can vary by several orders of magnitude depending on the quantum technology being used. 
It is only with the right combination of reliable and fast enough quantum hardware, quantum error correction, and quantum software that we will be able to achieve a quantum advantage over classical computing~\cite{hype}.

To reliably and efficiently estimate the resources required to execute a quantum computation we need automatic software tools.
Multiple quantum software packages exist that approach this problem from different angles.
Recently released resource estimation tools include Zapata AI's BenchQ~\cite{BenchQ}, Google's QUALTRAN~\cite{Qualtran}, and MIT Lincoln Lab's pyLIQTR~\cite{pyLIQTR}.

\begin{figure*}
    \centering
    \includegraphics[width=0.96\textwidth]{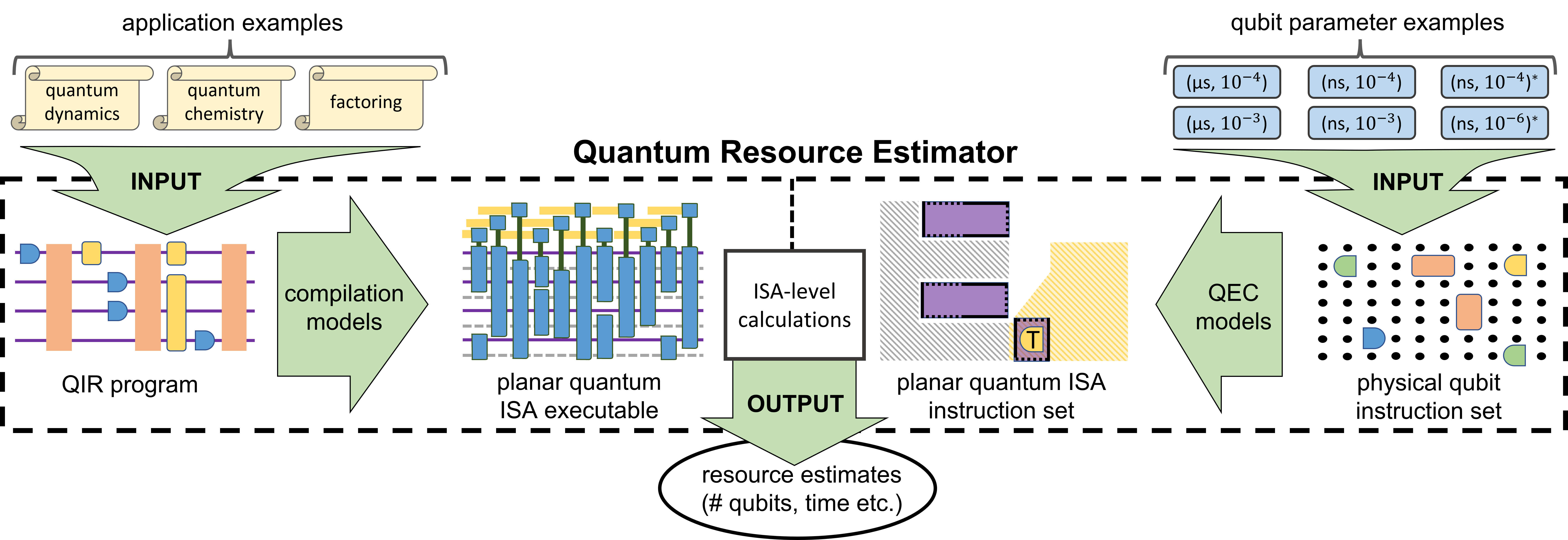}
	\caption{
    Figure from \cite{Beverland2022}, showing how the resource estimation tool uses the planar, quantum Instruction-Set Architecture (ISA) to process its inputs: a high level description of a quantum program and the low level physical qubit parameters. 
    The tool outputs resource estimates such as the number of physical qubits and time required to run the application on a machine with the specified hardware profile using the specified quantum error correction scheme.
    }
    \label{fig:QRE}
\end{figure*}

Azure Quantum Resource Estimator, described in the current paper, performs resource estimates on multiple levels, from the logical level (numbers of logical qubits, gates, and measurements, as well as logical circuit depth) down to the physical level. 
It takes into account the necessities of performing large-scale quantum error correction to implement the specific quantum algorithm, and is therefore able to calculate required quantities such as the code distance, the number of T state factories, and the number of physical qubits. 
The next section explains these capabilities further. 

\section{Resource estimation algorithm\label{sec:algorithm}}

Azure Quantum Resource Estimator (``the tool'') assumes the planar quantum Instruction-Set Architecture (ISA) as the logical abstraction level where the algorithm specification and the physical parameters of a chosen quantum hardware profile meet.
See Appendix~B in \cite{Beverland2022} for a detailed explanation of this abstraction. 
The tool enables the user to provide a high-level specification of an algorithm, which then gets automatically translated to the quantum ISA level. 
Section~\ref{sec:formats} gives more details on this. 

At the lower end of the stack, the tool enables the user to specify the physical error rates and durations of the primitive operations, the quantum error correction (QEC) scheme that is being used, some constraints on the component level, and the error budget for the algorithm success rate.
Section~\ref{sec:hardware_profiles} gives more details and examples of this specification. 

\begin{figure}
    \centering
    \includegraphics[width=0.48\textwidth]{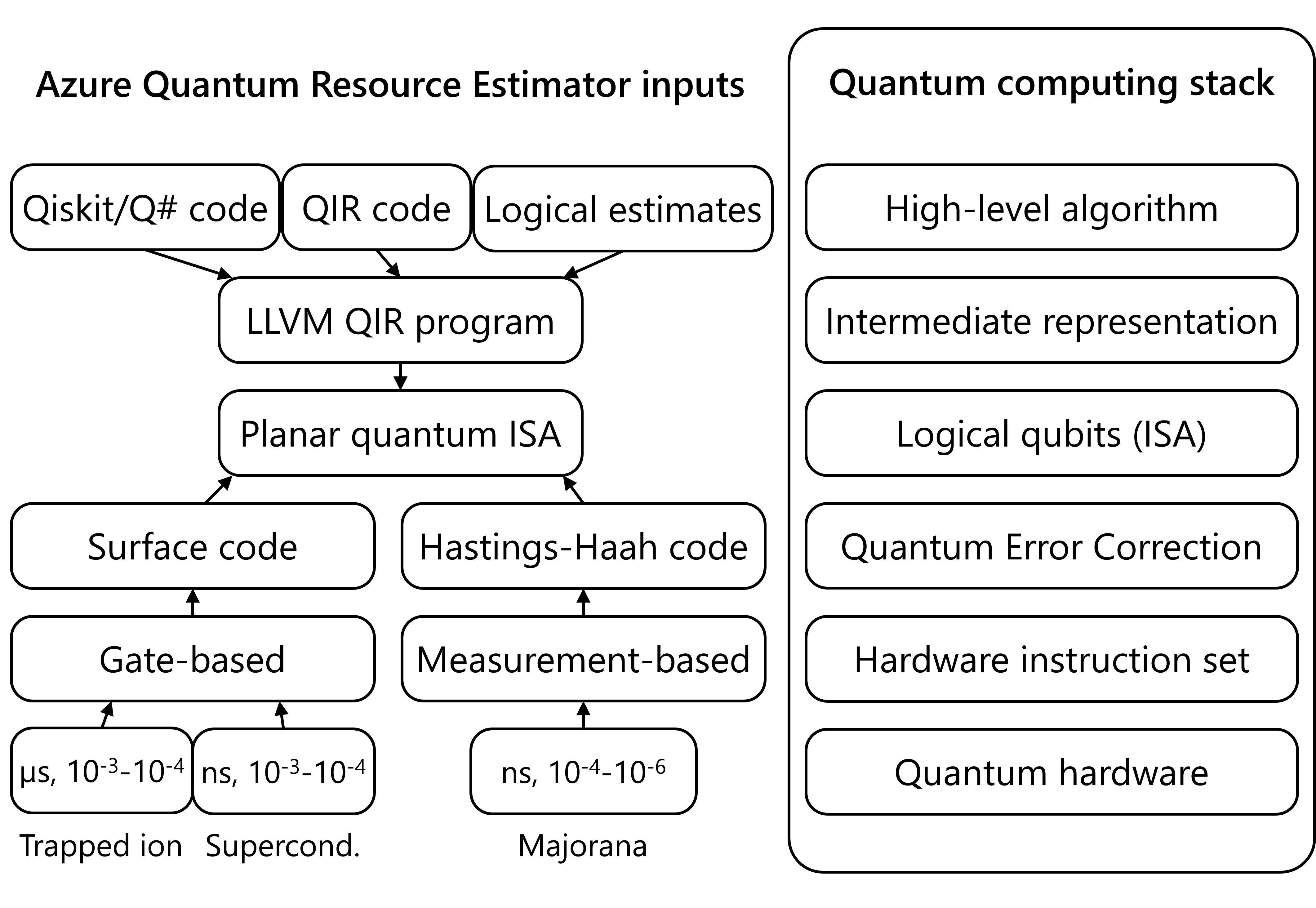}
	\caption{The inputs of Azure Quantum Resource Estimator on each level of the quantum stack.} \label{fig:stack_layers}
\end{figure}

The resource estimation procedure consists of multiple steps, following the flow of quantum program compilation and estimation from the logical layer down to the physical layer. 
Figure~\ref{fig:QRE} offers a high-level overview of the inputs and outputs of the algorithm, as well as the intermediate representations used on various stages of calculations.
Figure~\ref{fig:stack_layers} shows the inputs to the resource estimator tool and their correspondence to the layers of the quantum computing stack.
The specification of the algorithm can be provided in several different formats, such as a program in a high-level quantum language, low-level Quantum Intermediate Representation (QIR) code~\cite{QIRspec}, or prior logical estimates. 
Sections~\ref{section:pre-layout} through \ref{section:rqops} give an informal walk-through of the steps involved. 

\subsection{Pre-layout resource estimation}\label{section:pre-layout}

The first step of resource estimation is performed on the logical level before the layout step. 
The tool converts the input specification of the algorithm from a high-level programming language to QIR, and then uses the QIR program to calculate the number of logical qubits allocated by it (the ``width'' of the circuit), as well as the number of explicitly invoked T gates, single-qubit rotation gates, CCZ and CCiX gates, and measurements. 
If the input algorithm is specified in terms of prior logical estimates, then this step uses these logical estimates of the gate numbers instead.

\subsection{Algorithmic logical resource estimation}

The second step extends the logical resource estimates with several more parameters: the post-layout number of logical qubits, the rotation depth, the algorithmic depth, and the number of required T states. 

\subsubsection{The number of logical qubits after layout.}
Resource Estimator assumes 2D nearest-neighbor connectivity of the qubits, which is a common restriction in planar device platforms. 
To enable the generally required all-to-all connectivity of a quantum program, the tool uses additional logical qubits, alternating rows of algorithmic qubits with rows of auxiliary qubits used to implement multi-qubit Pauli measurements. 
Note that the tool does not (yet) analyze the qubit connectivity used in the algorithm to optimize the number of additional logical qubits at this step.

\subsubsection{Rotation depth of the algorithm.} 
Rotation depth is defined as the number of non-Clifford layers of gates in which each layer contains at least one arbitrary rotation gate.

\subsubsection{Algorithmic logical depth of the algorithm.}
Logical depth is defined as the number of logical cycles required to run the algorithm. 
As it uses QEC, the algorithm is executed via a sequence of multi-qubit measurements, hence the logical depth of the algorithm is expressed in terms of these measurements, which is the sum of the following numbers:
\begin{itemize}
    \item $1 \times$ (number of single-qubit measurements + number of single-qubit rotation gates + number of T gates) +
    \item $3 \times$ (number of CCZ gates + number of CCiX gates) +
    \item (number of T states used per rotation) $\times$ (rotation depth).
\end{itemize}

\subsubsection{The total number of required T states.}
This quantity is calculated as the sum of the following numbers:
\begin{itemize}
    \item $1 \times$ (number of T gates used in the code explicitly) +
    \item $4 \times$ (number of CCZ gates + number of CCiX gates) +
    \item (a multiplier that depends on the error probability of rotation synthesis) $\times$ (number of arbitrary rotation gates).
\end{itemize}

\subsection{Error correction}

The third step converts the logical estimates into physical estimates, taking into account the overhead introduced by error correction. 
The error correction scheme specified in the inputs of the tool, combined with the error budget for the algorithm, allows the tool to calculate the necessary and sufficient code distance of the QEC protocol used. 
Using this distance, it can then give the number of physical qubits per logical qubit and the runtime in nanoseconds of one logical cycle. 
The inverse of this runtime is referred to as the logical clock speed or rate of the computation. 
These parameters are then used to calculate the number of physical algorithmic qubits and the runtime in nanoseconds required to run the algorithm.

\subsection{T factories physical estimation}

This step estimates the physical resources required to produce the T states used during the algorithm execution. 
Based on the total number of required T states and the parameters of individual T factories specified as part of the tool input, it evaluates the number of T factory copies that will be running in parallel and the number of times each of them can be invoked during the runtime of the algorithm.
The input parameters allow the user to tweak the choices done on this step, trading off the number of qubits required to run these T factory copies for the total runtime of the T factories.
The number of physical qubits used to run T factories and the runtime of the T factories are the final components that contribute to the total physical resources required by the algorithm.

\subsection{rQOPS: Reliable Quantum Operations Per Second}\label{section:rqops}

By combining the results of the previous steps, the tool calculates the physical qubit resources that are needed to implement the specific quantum algorithm and its execution time in nanoseconds on the chosen hardware profile. 

To summarize the estimated quantum computational performance, the tool also expresses the \emph{reliable Quantum Operations Per Second (rQOPS)} rate of the computation~\cite{rQOPS-arxiv, rQOPS-Nayak}. 
This value, which is somewhat akin to the floating point operations per second (FLOPS) rate of classical computing, is estimated by the number of logical qubits multiplied by the logical clock speed (in Hertz).  
Depending on the hardware profile, the code distance, and the number of qubits, the rates for practical quantum solutions will typically sit between $\num{e2}$ rQOPS and $\num{e9}$ rQOPS.  

For more details on the steps of resource estimation algorithm, the conversion formulas used for the computations, and the assumptions made on each step see \cite{Beverland2022}.

\section{Azure Quantum Resource Estimator \label{sec:AQRE}}
\subsection{Azure Quantum Development Kit and Azure Quantum}

Azure Quantum Development Kit (\AQDK) is a collection of quantum software development tools, including the domain-specific programming language Q\#, sparse simulator, and Azure Quantum Resource Estimator (\AQRE)---a tool that performs automatic resource estimates for Q\# programs.

Azure Quantum is a cloud-based platform by Microsoft that provides access to a diverse set of quantum hardware from multiple partners and software services, including noisy and noiseless quantum simulators and Azure Quantum Resource Estimator service for Qiskit and QIR programs. 
When connected to an Azure Quantum workspace, the tool will act like a cloud target to which one can submit a resource estimation job, much like how one can submit jobs to other targets. 
After retrieving the output of this job, the results can further be parsed by the user using Python code. 

Several example Q\# projects and Jupyter notebooks that use Azure Quantum Resource Estimator are available~\cite{AQREnotebooks}, as is the detailed documentation of the tool~\cite{AQREdocs}.

\subsection{Algorithm specification formats}\label{sec:formats}
The quantum algorithm that is the input to \AQRE can be specified using a high-level programming language, a low-level intermediate representation, or using known logical estimates of the resources required. 
Examples that demonstrate these capabilities are available at \cite{AQREnotebooks}.

\subsubsection{Qiskit or Q\# code}
The easiest way to use Resource Estimator is to provide a quantum program in Q\# or Qiskit as the input. 
The tool compiles the program into Quantum Intermediate Representation (QIR) to enable unified processing of inputs in multiple quantum programming languages. 
Once the code is compiled into QIR, the tool will go through the code and track qubit allocation, qubit release, gate application, and measurement events. 
Doing this yields the statistics about logical-level resource requirements of the program that will then be used to produce physical-level resource estimates, as was described in Section \ref{sec:algorithm}.

\subsubsection{Quantum Intermediate Representation (QIR)}
The tool is built on top of Quantum Intermediate Representation and can use it as an input algorithm specification, either in raw form or emitted using PyQIR or another QIR-generation tool. 
This can be convenient for, e.g., processing algorithms that were generated automatically from a classical description rather than implemented in a high-level quantum language~\cite{soeken2022automatic}.

\subsubsection{Known logical estimates} 
The tool provides a Q\# operation \texttt{AccountForEstimates} that allows the user to define a ``virtual'' operation that uses known logical estimates---specified numbers of certain gates, measurements, and auxiliary qubits. 
This technique can be convenient for incorporating logical resource estimates from prior manual calculations into the overall cost of a larger Q\# program and converting them into low-level physical resource estimates.

Alternatively, you can convert known logical estimates to detailed physical estimates without writing any Q\# code by using Python function \texttt{LogicalCounts}.

\subsection{Hardware specification parameters} \label{sec:hardware_profiles}
Resource Estimator allows to customize the resource estimation parameters to match the characteristics of the target hardware.
The most important parameters are the following. 

\subsubsection{Physical qubit parameters}
This group of parameters specifies the properties of the underlying physical qubits: the underlying instruction sets (gate-based or Majorana) and operation times and error rates for the instructions in those sets, as well as the error rate for the ``idle'' operation.
For gate-based qubits, the instruction set is characterized by  operation times and error rates for single- and two-qubit gates, T gates, and single-qubit measurements. 
For Majorana qubits, the instruction set is characterized by operation times and error rates for single- and two-qubit measurements and T gates. 

The default qubit models include parameter sets that roughly correspond to gate-based superconducting transmon qubits, gate-based trapped ion qubits, and Majorana qubits, with each model featuring two regimes corresponding to realistic and optimistic estimates of the physical error rates. 
It is also possible to customize a subset of the parameters of the default models while keeping others at their default values, or to specify a completely custom model.

\subsubsection{QEC scheme}
This group of parameters specifies the parameters of the error correction scheme assumed to be used by the hardware. 
The default QEC protocols include surface codes and floquet/Hastings-Haah codes. 

The error correction scheme is defined by two numeric parameters, crossing pre-factor and error correction threshold, and two formula parameters, logical cycle time and the number of physical qubits per logical qubit. 
The formula parameters are strings that use simple arithmetic operations and variables for one- and two-qubit gate and measurement times and the code distance (computed based on the physical qubit properties and the numeric parameters of the scheme).

It is also possible to customize some of the parameters of the default protocols while keeping the others at their default values, or to specify a completely custom protocol.

\subsubsection{Error budget}
This parameter specifies the overall error budget for the algorithm, i.e., the maximum failure rate allowed. 
It is also possible to specify separately the error budget to implement logical qubits, error budget to produce T states through distillation, and error budget to synthesize arbitrary rotation gates.
Error budget parameters inform the choice of the QEC parameters used in resource estimation.

\subsubsection{T factory constraints} 
These parameters allow to specify the maximum number of T factory copies and the logical cycle slowdown to optimize the qubit versus runtime trade-off. Slowing down the program execution allows to run fewer T factory copies in parallel, reducing the qubit requirements for T factories but possibly impacting the QEC parameters and the overall runtime.

\subsubsection{Distillation units} 
Resource Estimator allows to specify one of the predefined T factories distillation algorithms or to customize the distillation algorithms. A distillation algorithm specification includes the numbers of input and output T states, failure probability and output error rate (specified as formula strings), and specifications for the distillation unit when it is applied to physical and/or logical qubits.

\subsection{Output data of Azure Quantum Resource Estimator}
The output data of Azure Quantum Resource Estimator is broken down into the following groups.

\subsubsection{Physical resource estimates}
The main outputs of the tool are the runtime of the algorithm, the reliable quantum operations per second (rQOPS) rate, and the number of physical qubits required to implement the algorithm. 

\subsubsection{Resource estimates breakdown}
This output group includes more detailed estimates on the higher levels compared to the physical qubits level, such as the number of logical qubits used by the algorithm, its logical depth, and the parameters of T factories required to run it.

\subsubsection{Logical qubit parameters}
This group includes the details of the QEC scheme necessary to run the algorithm within the given error budget.

\subsubsection{T factory parameters}
This group provides detailed parameters of T factories necessary to run the algorithm.

\subsubsection{Pre-layout logical resources}
This group includes logical-level resource estimates: the number of logical qubits used by the algorithms and numbers of logical gates, including T, CCZ, CCiX and rotation gates.

\subsubsection{Assumed error budget}
This group describes the breakdown of the total error budget into the logical, the T distillation, and the rotation synthesis error probability. 

\subsubsection{Physical qubit parameters}
Contains the assumed properties of qubits and elementary operations on them. 

\subsubsection{Assumptions}
The last group describes the overall assumptions of the estimation process. 

\begin{figure*}
    \centering
    \includegraphics[width=0.96\textwidth]{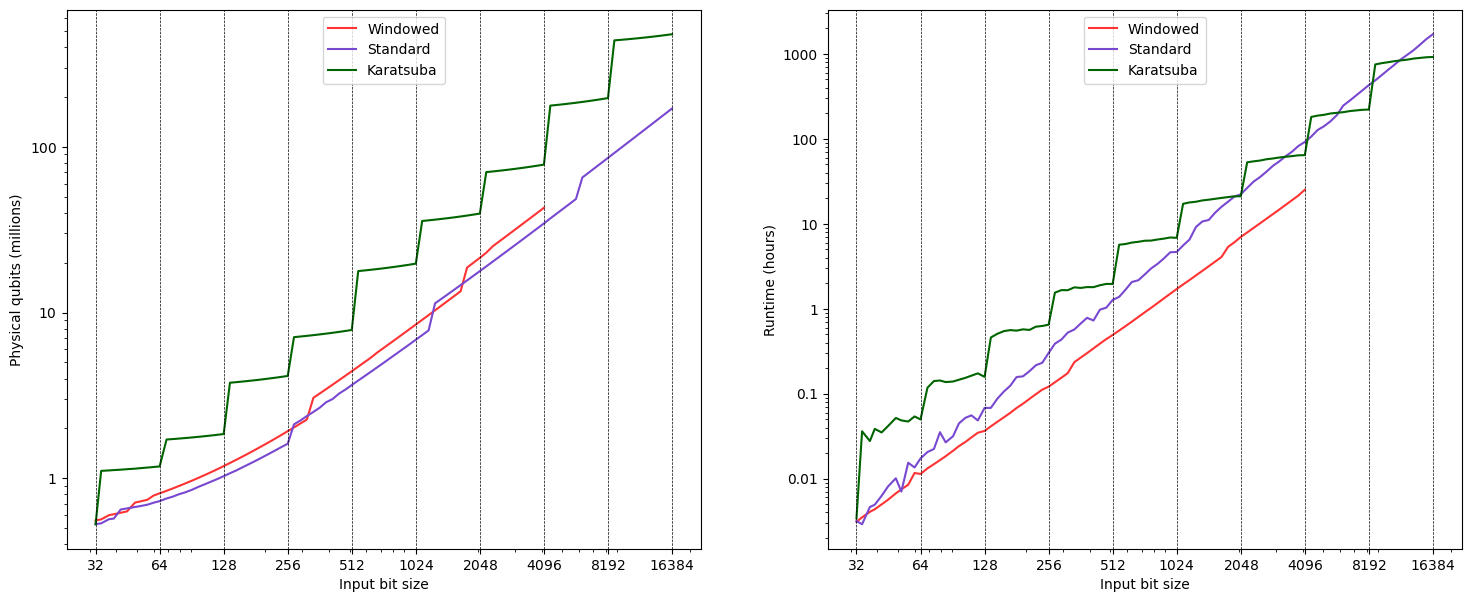}
    \caption{The number of physical qubits and total runtime for the three multiplication algorithms, estimated for input sizes from $\num{32}$ bits to $\num{16384}$ bits. 
    The results were produced for hardware profile \texttt{qubit\_maj\_ns\_e4} with floquet code QEC scheme and total error budget $\num{e-4}$. 
    At these parameters, the QEC code distance varies from distance $\num{9}$ (at $\num{32}$ bits) to distance $\num{17}$ (at $\num{16384}$ bits). 
    The increments in code distance can be seen by the upward jumps in the physical qubits plot for windowed and standard multiplication algorithms. At distance $\num{2048}$ bits a distance $\num{15}$ code is used.}
    \label{fig:increasing_bitsizes}
\end{figure*}

\section{Integer multiplication use case\label{sec:use case}}
As an example of using the tool in practice, we conclude by showing the resource estimates for three quantum algorithms for large integer multiplication: the standard \emph{long multiplication}, \emph{Karatsuba multiplication,} and \emph{windowed multiplication.}
For $n$ bit numbers the classical computational complexity of long multiplication is $\Omega(n^2)$, whereas Karatsuba multiplication has improved complexity $O(n^{\log_2 3})$ ($\log_2 3 \approx 1.58$). 
Karatsuba algorithm implementation using quantum circuits is described in \cite{gidney2019asymptotically}. 
The third, windowed algorithm tries to improve on the time complexity of the standard algorithm by using the quantum circuit equivalent of a look-up table, which is described in \cite{gidney2019windowed}. 
For our analysis of these algorithms we use the Q\# implementations from \cite{gidney2019asymptotically} and \cite{gidney2019windowed}.
% Note that the first two algorithms multiply two quantum registers \cite{gidney2019asymptotically}, while the third algorithm multiplies a single quantum register by a constant\cite{gidney2019windowed}. 
This data is based on a recent work by Hansen, Joshi, and Rarick~\cite{AQET-RE}.

\begin{figure}
    \centering
    \includegraphics[width=0.48\textwidth]{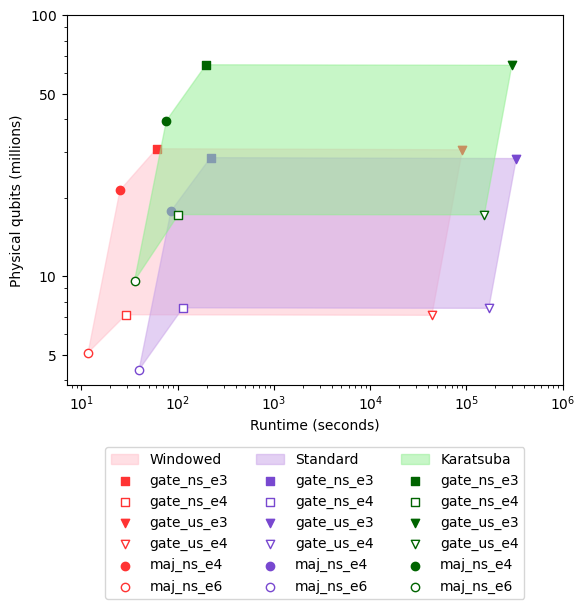}
	\caption{The number of physical qubits and algorithm runtime across six different hardware profiles for the three multiplication algorithms, estimated for $\num{2048}$-bit integers. 
	The hardware profiles are described in Section~III.C of \cite{Beverland2022}. The results are produced for total error budget $\num{0.0001}$. The estimates for gate-based hardware profiles used the surface code QEC scheme with default parameters, and Majorana hardware profiles used floquet code QEC scheme with default parameters.} \label{fig:platform_comparison}
\end{figure}

Figure~\ref{fig:increasing_bitsizes} shows the resource estimation results of the three algorithms for multiplying integers of increasing bit sizes $n$ with total error budget  $\num{e-4}$, and assuming the hardware profile \texttt{qubit\_maj\_ns\_e4} with parameters:
\begin{itemize}
\item{gate operation time: $\qty{100}{\ns}$}
\item{measurement operation time: $\qty{100}{\ns}$}
\item{Clifford error rate: $\num{e-4}$}
\item{non-Clifford error rate: $\num{0.05}$}
\item{QEC scheme: Hastings-Haah}
\end{itemize}
See \cite[Figure II and Table V]{Beverland2022} for more details on this profile. 

Supplementing this,  Figure~\ref{fig:platform_comparison} shows the resource estimation results of the three algorithms for multiplying $\num{2048}$-bit integers across six default hardware profiles~\cite{Beverland2022}.

% Updated number for Karatsuba
From this analysis we can see, for example, that the Karatsuba algorithm requires more physical qubits than the other two algorithms. 
With the respect to the runtime, the Karatsuba algorithm first shows a runtime improvement over the standard multiplication around $\num{4096}$-bit inputs, and it does not become consistently faster than standard multiplication until inputs size over $\num{16384}$-bits. 
We can also see that for $\num{2048}$-bit numbers, the windowed algorithm uses $\num{1.12e11}$ logical quantum operations and $\num{20597}$ logical qubits. 
The estimated runtime varies between $\num{12}$ and $\num{9e4}$ seconds (depending on the hardware profile), hence the subroutine computes at between $\num{1.37e6}$ and $\num{9.1e9}$ rQOPS.  

Some conclusions that we can draw from this is that the asymptotic superiority of the Karatsuba algorithm does not materialize for realistic input sizes and that this classically trivial task requires significant quantum resources.   
These conclusions are both significant and somewhat unexpected, thus underlining the importance of performing resources estimation to assess the feasibility of quantum solutions.

\section*{Acknowledgements}
% \balance % this is a bugy ACM command that should be entered in the left column of the last page. 

We thank Azure Quantum team for developing Azure Quantum Resource Estimator.
We thank Ethan Hansen, Sanskriti Joshi, and Hannah Rarick for their prior collaboration on estimating resources for multiplication algorithms as part of their project within the UW EE522: Quantum Information Practicum class at the University of Washington.


\bibliographystyle{IEEEtran}
\bibliography{bib}

% \appendix

\end{document}